# NASA's astonishing evidence that c is not constant: The pioneer anomaly

E. D. Greaves
*Universidad Simón Bolívar, Laboratorio de Física Nuclear, Apartado 89000, Caracas 1080 A, Venezuela. E-mail: egreaves20002000@yahoo.com*

For over 20 years NASA has struggled to find an explanation to the Pioneer anomaly. Now it becomes clear the solution to the riddle is that they have uncovered evidence that c, the speed of light, is not quite a universal constant. Using J. C. Cure's hypothesis that the index of refraction is a function of the gravitational energy density of space and straightforward Newtonian mechanics, NASA's measurements provide compelling evidence that the speed of light depends on the inverse of the square root of the gravitational energy density of space. The magnitude of the Pioneer anomalous acceleration leads to the value of the primordial energy density of space due to faraway stars and galaxies: $1.0838 \times 10^{15}$ Joule/m$^3$. A value which almost miraculously coincides with the same quantity: $1.09429 \times 10^{15}$ Joule/m$^3$ derived by J. C. Cure from a completely different phenomenon: the bending of starlight during solar eclipses.

*Introduction*
Anderson and collaborators at the Jet Propulsion Laboratory (JPL) have reported [1] an apparent, weak, long range anomalous acceleration of the Pioneer 10 and 11 with supporting data from Galileo, and Ulysses spacecraft. [2, 3] Careful analysis of the Doppler signals from both spacecraft have shown the presence of an unmodeled acceleration towards the sun. By 1998 it was concluded from the analysis, that the unmodeled acceleration towards the Sun was $(8.09 +/- 0.20) \times 10^{-10}$ m/s$^2$ for Pioneer 10 and of $(8.56 +/- 0.15) \times 10^{-10}$ m/s$^2$ for Pioneer 11. In a search for an explanation, the motions of two other spacecraft were analyzed: Galileo in its Earth-Jupiter mission phase and Ulysses in a Jupiter-perihelion cruise out of the plane of the ecliptic. It was concluded that Ulysses was subjected to an unmodeled acceleration towards the Sun of $(12 +/- 3) \times 10^{-10}$ m/s$^2$. To investigate this, an independent analysis was performed of the raw data using the Aerospace Corporation's Compact High Accuracy Satellite Motion Program (CHASMP), which was developed independently of JPL. The CHASMP analysis of Pioneer 10 data also showed an unmodeled acceleration in a direction along the radial toward the Sun. The value is $(8.65 +/- 0.03) \times 10^{-10}$ m/s$^2$, agreeing with JPL's result. Aerospace's analysis of Galileo Doppler data resulted in a determination for an unmodeled acceleration in a direction along the radial toward the Sun of, $(8 +/- 3) \times 10^{-10}$ m/s$^2$, a value similar to that from Pioneer 10. All attempts at explanation of the unmodeled acceleration as the result of hardware or software problems at the spacecraft or at the tracking stations have failed. A very detailed description of the Pioneer anomaly, of the measurements and of the analysis was given by the JPL team [4]. Two conferences have been carried out on the subject, in 2004 [5] and in 2005 [6] and although several explanations have been advanced, no clear consensus exists of the cause of the weak [7] anomalous acceleration experienced by the various spacecraft. With no plausible explanation so far, the possibility that the origin of the anomalous signal is new physics has arisen.[8] Very recently evidence of the puzzling phenomenon was found in the motion of other spacecraft. [9]

The Pioneer anomalous acceleration $a$ is derived from the Doppler drift $\Delta f$ of the base frequency $f_o$ detected: $\Delta f = f_o \left( a/c \right)$ In this paper the anomalous drift is shown to be due to a change of the index of refraction of vacuum, a function of the gravitational energy density of space predicted by the Curé hypothesis [10]. It affects $c$ the speed of light in space far from the influence of the sun.

**1.- Energy density of space.**
By energy density of space we mean the classical energy density (Energy per unit volume) associated with the potential energy of all forms of force: electric, magnetic, gravitational or any other force in nature. In particular, to be associated to gravitational energy of nearby massive bodies such as the Sun and the Earth which we can readily calculate, and to the gravitational energy density produced by the gravitational field of the stars and far away galaxies, not so easily estimated.

The energy density of space associated with the presence of static electric $E$ and magnetic $B$ fields are given by:

$$\rho = \frac{1}{2}\varepsilon_o E^2 + \frac{1}{2\mu_o} B^2 \qquad (1)$$

Where $\varepsilon_0$ and $\mu_0$ are the electric permittivity and the magnetic permeability of space respectively. The equivalent energy density associated with a gravitational field $g$ (m/s$^2$) is given by

$$\rho = \frac{1}{2}\frac{g^2}{4\pi G} \qquad (2)$$

with $G$ the Universal constant of gravitation. Hence any volume of space is immersed in the universal primordial field of energy $\rho^*$ which includes the immediate gravitational field due to the presence of our own galaxy superimposed on the energy fields of all far-away galaxies. Thus the energy density in the surface of the Earth and in the proximity of the Sun is given by:

$$\rho = \rho^* + \rho_S + \rho_E \qquad (3)$$

where the energy density due to the Sun $\rho_S$ produced by the gravitational effect of the mass of the Sun $M_S$ is obtained from (2) with $g = GM_S/r^2$

$$\rho_S = \frac{GM_S^2}{8\pi r^4} \qquad (4)$$

Here $r$ is the distance from the centre of the Sun to the point in question. And $\rho_E$ is the energy density due to the gravitational effect produced by the mass of the Earth and is calculated in analogous manner. The acceleration of gravity $g_S$ due to the Sun at the radius of the Earth's orbit is $g_S = 0.00593$ m s$^{-2}$. Hence the Sun's energy density at the Earth orbit is $\rho_S = 2.097 \times 10^4$ Joules/m$^3$. With the Earth's acceleration of gravity the energy density due to the Earth at the surface is $\rho_E = 5.726\ 10^{+10}$ Joules/m$^3$ and the universal primordial energy density $\rho^*$ is estimated [10] at 1.09429 x $10^{15}$ Joules/m$^3$. This is a value arrived at by an analysis of the deflection of light by the Sun's energy field considered as a refraction phenomenon as reviewed below. [11]

J.C. Curé [10, p. 276] explains the energy density of space in the following illuminating words:

> "Every celestial body is surrounded by an invisible envelope of gravitostatic energy caused by the matter of the body and given by Eq. (104). (Our Eq. 4) To proceed with a colorful description, let us assign a yellow color to the sun's gravitostatic energy. Let us picture the background cosmic energy with a bluish color. Now we can see, in our imagination, that the sun is surrounded with a green atmosphere of energy. The green color fades away into a bluish color as we recede from the sun."

**2.- Effect of energy density of space**

Now let us consider the hypothesis that the speed of light is a function of the energy density of space $\rho$ which in the neighbourhood of the sun is determined by a constant background value due to the distant galaxies plus a smaller value due to the gravitational presence of the Sun's mass as seen by (3) above.

We assume the speed is inversely proportional to the square root of the energy density by the use of the Curé hypothesis [10, p 173] given by relation (5):

$$c' = \frac{k}{\sqrt{\rho^* + \rho_S + \rho_E}} \qquad (5)$$

This implies that the speed of light decreases near the Sun and increases far away from the sun. We may then assign an index of refraction $n$ to space such that $n = 1$ in vacuum space near the Earth, as we usually do, and assign an index $n' < 1$ far away from the Sun, in deep space, where the speed of light $c'$ is greater and is given by:

$$c' = \frac{c}{n'} \qquad (6)$$

so that the index of refraction there is $n' = c/c'$.

Using (5) we may write expressions for $c$ and $c'$ and obtain the index of refraction, $n'$, far away from the Sun in terms of $\rho_{S1AU}$ the energy density of the Sun at the distance of the Earth: one Astronomical Unit ($r$ = 1 AU), $\rho_E$ the energy density of the Earth at the surface, $\rho_{Sfar}$, the energy density of the Sun, relatively far away but in the vicinity of the Sun and $\rho^*$ the interstellar primordial energy density in the vicinity of the Sun [12] as:

$$n' = \frac{\sqrt{\rho^* + \rho_{Sfar} + \rho_{Efar}}}{\sqrt{\rho^* + \rho_{S1AU} + \rho_E}} \qquad (7)$$

Strictly speaking, relation (7) should contain in the numerator and denominator the gravitational energy density due to all the other planets. However, the contribution is negligible due to the $1/r^4$ factor in the energy density, unless $n'$ is being calculated near a planet.

At this point it is fitting to address the order of magnitude of the quantities being discussed. With $n = 1$ at the Earth at 1 AU from the Sun, the index of refraction $n'$ further away from the Sun is dependent on the relative magnitudes of the energy density values that enter into Eq. (7), i.e. the relative value of the Sun's energy density, the Earth's energy density and the primordial energy density $\rho^*$ of space due to the stars and far-away galaxies.

If we plot relation (4) we find that $\rho_S$ falls of rapidly as we go away from the Sun, see Fig. 1, and it becomes negligible for distances of say $r > 10$ AU compared to the universal primordial energy density estimated by Curé [10, p 279] at $1.094291 \times 10^{15}$ Joules m$^{-3}$. Entering these values into (7) we find that $n'$ is smaller than one for $r > 1$ AU, and it is also smaller than one for $r < 1$ AU due to the energy density of the Earth which, near the surface, is much greater that the sun's energy density. But the numerical value of $n'$ is very nearly one, differing only by a very small amount (see Table I). Hence the values of the speed of light calculated at different distances from the Sun changes little from the accurately measured value on the surface of the Earth at a distance of 1 AU from the Sun. These minute changes in the speed of light or of the index of refraction of space are consistent with the tiny magnitudes of the accelerations reported by the Pioneer anomaly. With our knowledge of the energy density of the Sun and Earth, relation (7) for the index of refraction $n'$ may be used to determine the primordial energy density of space, $\rho^*$, if we do an independent measurement of the index of refraction of space, $n'$, far away from the Sun. Solving for $\rho^*$ we get:

$$\rho^* = \frac{\rho_{Sfar} + \rho_{Efar} - n'^2 (\rho_{S1AU} + \rho_E)}{n'^2 - 1} \tag{8}$$

In this relation $n'$ is the index of refraction at the distance where $(\rho_{Sfar} + \rho_{Efar})$ is calculated.

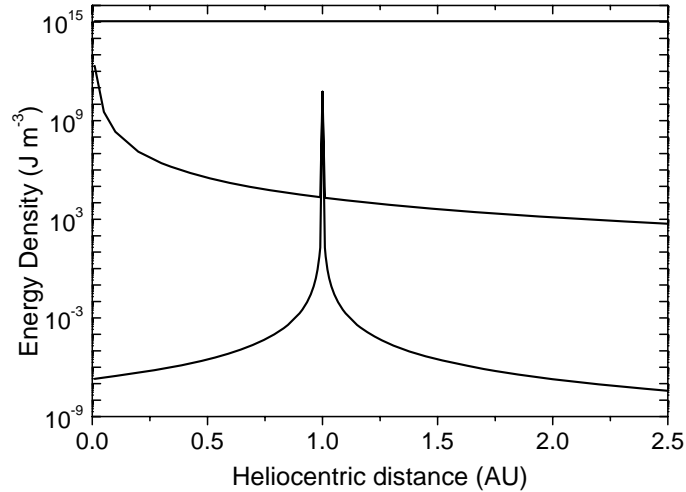

Figure 1. Energy density of space as a function of distance from the sun. Top line, energy density due to the stars. Middle line, Sun's gravity + Earth. Bottom line, energy density due to Earth. (Along a radial line Sun–Earth)

**3.- Doppler Effect.**

The frequencies of signals received from spacecraft are affected by their movement through the Doppler Effect. In fact the first order Doppler Effect is normally used to determine the speed of distant spacecraft. An accurate oscillator "clock" on board emits a signal in the form of an electromagnetic wave at a base frequency $f_o$. If the spacecraft moves at a velocity, $v$, relative to the receiving station the frequency $f$ of the clock as perceived by the receiver is shifted from $f_o$ by an amount $\Delta f$:

$$\Delta f = f_o - f = f_o \left(\frac{v}{c}\right) \tag{9}$$

Hence

$$f = f_o - f_o \left(\frac{v}{c}\right) = f_o \left(1 - \frac{v}{c}\right) \tag{10}$$

This is the frequency received when $v$ is in the direction away from the receiver, i. e. the signal of a receding spacecraft is Doppler-shifted towards lower frequencies (red-shifted). The reverse occurs if the spacecraft moves towards the receiver, in which case the received signal is Doppler-shifted towards higher frequencies (blue-shifted).

Above we assumed a "clock" on board for clarity in the argument. However, in the case of the Pioneer spacecrafts this is not true. The signals transmitted by the Pioneer spacecrafts are re-transmission of Earth-sent signals. Assume the frequency transmitted from Earth is $f_e$, the spacecraft is in motion relative to Earth hence the frequency of the signal received at the spacecraft for retransmission is not $f_e$ but rather a Doppler shifted frequency $f_o$. The shift is given by a relation analogous to (9): In the spacecraft frame of reference Earth is receding with speed v. Hence the signal received is Doppler shifted by an amount $\Delta f_s$

$$\Delta f_s = f_e - f_o = f_e \left(\frac{v}{c}\right)$$

Solving for $f_o$ we obtain a relation like (10):

$$f_o = f_e \left(1 - \frac{v}{c}\right)$$

Which substituted in (10) gives:

$$f = f_e \left(1 - \frac{v}{c}\right)^2 = f_e \left(1 - \frac{2v}{c} + \frac{v^2}{c^2}\right)$$

Neglecting the second order term the Doppler-shifted frequency $f$ received on Earth due to the spacecraft in motion with speed $v$ is

$$f = f_e \left(1 - \frac{2v}{c}\right)$$

and the change relative to the Earth-sent frequency is:

$$\Delta f = f_e - f = f_e \left(\frac{2v}{c}\right) \tag{11}$$

**4.- Effect of Gravity on speed of spacecraft**

A spacecraft that is receding into deep space away from the Sun does not move with a constant velocity. This is because it is affected by the gravitational attraction of the Sun.

The effect is that the receding spacecraft is affected by a change of speed towards the Sun which is equal to the magnitude of the Sun's acceleration of gravity at the position of the spacecraft. The acceleration is in the direction of the Sun which is approximately in a direction opposite to its receding speed.

For a deep space probe spacecraft the acceleration $a$ is given from Newton's second law by $a = F/m$ with $F$ the gravitational force of the Sun on the spacecraft and $m$ the spacecraft mass. $F$ is given by Newton's relation: $F = GM_S m/r^2$ with $G$ the universal constant of gravitation, $6.67300 \times 10^{-11}$ m$^3$ kg$^{-1}$ s$^{-2}$, and $M_S$ the Sun's mass, $1.98892 \times 10^{30}$ Kg, hence the acceleration of the spacecraft is:

$$a = \frac{GM_s}{r_s^2} \tag{12}$$

where $r_s$ is the radial distance from the spacecraft to the centre of the Sun.

The speed of the spacecraft is time dependent and is given by: $v = v_0 - at$ with $v_0$ the speed at some time $t = 0$, and $a$ the acceleration given by (12):

$$v = v_o - \left(\frac{GM_s}{r_s^2}\right)t \tag{13a}$$

If we wish to take into account the gravitational force of the Earth we must include a term similar to (12):

$$v = v_o - \left(\frac{GM_s}{r_s^2}\right)t - \left(\frac{GM_e}{r_e^2}\right)t \tag{13b}$$

Where $M_e$ is the mass of the Earth $5.98 \times 10^{24}$ Kg and $r_e$ is the distance to the spacecraft from the centre of the Earth.

**5.- Doppler effect with $c$ affected by the energy density of space**

Let us now consider a Pioneer spacecraft far in deep space, in a region of space where *n' < 1* re-transmitting an Earth-sent base frequency $f_e$ and moving away from a receiver station at a hypothetical *steady (constant) velocity $v$*.

The frequency $f$ and the frequency shift $\Delta f$ of the signal perceived by a receiver will not be given by relation (11) above but rather by:

$$\Delta f' = f_e - f' = f_e\left(\frac{2v}{c'}\right) \tag{14}$$

The primed variables are the values affected by the fact that the speed of light $c'$ in the region of the spacecraft is different.

Substituting $c' = c/n'$ we get:

$$\Delta f' = f_e - f' = f_e\left(\frac{2v}{c}\right)n' \tag{15}$$

The meaning of Eq. (15) is that the frequency shift perceived at the receiving station is smaller because *n' < 1*. Accordingly it would correspond to a smaller Doppler shift and hence interpreted by an observer, unaware of the value of *n'*, as due to a receding velocity of the spacecraft that is *smaller* that it actually is.

Now let us consider the effect on the Doppler signals on a spacecraft whose speed is affected by the gravitational attraction of the Sun and the Earth. The speed in not constant but rather a function of time given by Eq. (13) above, hence it is Doppler shifted by:

$$\Delta f' = 2 f_e \frac{(v_o - \frac{GM_s t}{r_s^2} - \frac{GM_e t}{r_e^2})}{c} n'$$

With $\Delta f'$ being a function of time, the time rate of change of the Doppler shifted signal is:

$$\frac{d\Delta f'}{dt} = -\frac{2 f_e n' G}{c} \left( \frac{M_s}{r_s^2} + \frac{M_e}{r_e^2} \right) \quad (16a)$$

However, if we neglect the change of the speed of light due to the energy density of space we would have the previous relation with $n' = 1$ as follows:

$$\frac{d\Delta f}{dt} = -\frac{2 f_e G}{c} \left( \frac{M_s}{r_s^2} + \frac{M_e}{r_e^2} \right) \quad (16b)$$

Hence the "Excess" Doppler shift $E_D$ (Hz/s) due to the effect of the energy density of space is given by the difference between these two relations:

$$E_D = \frac{d\Delta f'}{dt} - \frac{d\Delta f}{dt}$$

Or

$$E_D = \frac{2 f_e G}{c} \left( \frac{M_s}{r_s^2} + \frac{M_e}{r_e^2} \right) (1 - n') \qquad \text{(Hz/s)} \quad (17)$$

Relation (17) gives the "Excess" Doppler signal that is detected by a receiving station on Earth and interpreted as an anomalous acceleration towards the Sun due to the effect on the Doppler frequency by the higher speed of light in the interstellar medium as compared with the speed of light, $c$, on Earth.

Upon examination of Eq. (17) we see that the term in the parenthesis, $(1-n')$, is very small owing to the fact that $n'$ is smaller than one, but very near to one. At a distance of 20 AU from the sun this term is equal to 0.0000572. The term on the right of Eq. (17), excluding $(1-n')$, is the factor $(2 f_e / c)$ times the gravitational acceleration of the Sun and the Earth at the distance $r$, i.e. it is the drift of the Doppler signal due to the gravitational acceleration at that point. An acceleration which is mainly due to the Sun.

The Pioneer anomaly reported as a weak acceleration, $a$, toward the Sun is calculated from the time rate of change of the Doppler shift, Eq. (11):

$$E_D = \frac{d\Delta f}{dt} = \frac{2 f_e}{c} \frac{dv}{dt} = \frac{2 f_e}{c} a \qquad \text{(Hz/s)}$$

Hence the anomalous acceleration is:

$$a = G \left( \frac{M_s}{r_s^2} + \frac{M_e}{r_e^2} \right) (1 - n') \qquad \text{(m/s}^2\text{)} \quad (18)$$

With $n'$ given by Eq. (7).

Examination of (18) and (7) shows that the only unknown parameter is $\rho^*$, the primordial energy density of space due to the stars and far-away galaxies. Hence we are able to predict the magnitude of the Pioneer anomaly with $\rho^*$ as a single adjustable parameter.

We may use Eq. (18) in several ways:

i.- With the empirical value of the "Excess" Doppler shift, $E_D$, measured by Anderson and collaborators we can calculate what is the index of refraction $n'$ for a particular position of the deep space probes. Solving Eq. (17) for $n'$:

$$n' = 1 - \frac{E_D c}{2 f_o G \left( \frac{M_s}{r_s^2} + \frac{M_e}{r_e^2} \right)} \tag{19}$$

This then allows determination of the speed of light $c'$ in that position with $c' = c/n'$. It also allows calculation of the energy density of space $\rho^*$ due to the primordial energy field with relation (8) assuming the Curé hypothesis given by relation (5).

ii.- The second way we can use Eq. (18) is to calculate independently the values of the unmodeled acceleration as a function of distance from the Sun which is what is reported [1,4]. Eq. (18) may be written in terms of the true acceleration of gravity $a_g$, Eq. (12), as:

$$E_D = \frac{2 f_e a_g}{c}(1 - n')$$

Hence the "Excess" Doppler signal detected (Hz/s) is due to a fictitious "Excess" acceleration $E_a$ given by the real acceleration of gravity $a_g$ (m/s$^2$) times the factor $(1-n')$, i.e.

$$E_a = a_g (1 - n')$$

The factor $(1-n')$ is due to the variation of the index of refraction of space, or the change of the speed of light due to the energy density of space.

We wish to calculate this expression for the "Excess" acceleration $E_a$ as a function of the distance from Earth. We take into account only the effect of the Sun, due to its large mass, and of the Earth due to its large magnitude in its proximity, and neglect the effect of all the other planets. Using (7) and (12) in the previous relation the "Excess" acceleration is given by:

$$E_a = G\left(\frac{M_s}{r_{sx}^2} + \frac{M_e}{r_{ex}^2}\right)\left(1 - \sqrt{\frac{\rho^* + \frac{G}{8\pi}\left(\frac{M_s^2}{r_{sx}^4} + \frac{M_e^2}{r_{ex}^4}\right)}{\rho^* + \frac{G}{8\pi}\left(\frac{M_s^2}{r_{S1AU}^4} + \frac{M_e^2}{r_e^4}\right)}}\right) \tag{20}$$

Where $r_{Sx}$, $r_{ex}$ are the distances from the centre of the Sun and Earth to the $x$ position of the spacecraft, $r_{S1AU}$ and $r_e$ are the distances to the surface of the Earth, and $M_s$, $M_e$ are the masses of the Sun and Earth respectively.

## 6.- Results
Here we show the numerical results of calculations using the theory above.

i.- With the use of (18) and of data published [Ref. 4, p 15] of the frequency used in the transmission to the pioneer spacecraft of $f_e$ = 2295 MHz and the "Excess" Doppler shift, $E_D$, a steady frequency drift of $(5.99 \pm 0.01) \times 10^{-9}$ Hz/s from the Pioneer 10 spacecraft [4, p 20] we calculate that the index of refraction $n'$ at 20 AU from the Sun is:

$$n' = 0.9999735679^{\dagger} \qquad (21)$$

With this value the accepted speed of light measured on the Earth at 1 AU as $c$ = 299792458 m/s becomes at 20 AU the slightly higher value of $c'$ = 299800382 m/s .

The value (21) is the result of an empirical measurement of the index of refraction of space at 20 AU by NASA's careful measurements of Pioneer signals.

With this value and the use of Eq. (8) we can calculate the primordial energy density of space $\rho^*$, using the Sun's and the Earth's energy density at 1AU and at 20 AU. The value calculated is:

$$\rho^* = 1.0838. \times 10^{15} \text{ Joule/m}^3. \qquad (22)$$

This value coincides with the value of $\rho^* = 1.09429 \times 10^{15}$ Joule/m$^3$ calculated by Curé on the basis of an entirely different phenomenon: The bending of starlight rays by the gravitational field of the Sun. We outline here the calculation done by Jorge Céspedes-Curé [10, p. 279]. It consists of using the hypothesis of Eq. (5) interpreted as a change of the index of refraction of space, and using the analysis carried out by Prof. P. Merat [13] in 1974 [10, p 274] for 297 starlight deflections measured in 9 observations of 6 solar eclipses. With the results of Merat's analysis of the astronomical observations of the bending of starlight rays by the gravitational field of the Sun, Curé determines the energy density of space.

ii.- The result of calculating the Pioneer anomaly predicted by (20) as a function of distance is shown in Fig. 2. To construct this curve the space energy density $\rho^*$ was used as an adjustable parameter. The value chosen $\rho^* = 0.25 \times 10^{15}$ Joules/m$^3$ gives a better fit around 30 AU to the experimentally measured anomalous acceleration as a function of heliocentric distance.[14]

**7.- Discussion.**

The measurements of the Pioneer anomaly are not very precise. They are of the same order of magnitude of the errors in the measurements and with this imprecision they do not show a clear variation with the distance to the sun. However, considering the wildly different, magnitudes of the data that enter the relations used to calculate the energy density given by (22) (Gravitational constant, mass of the Sun and Earth, both masses squared, speed of light, distance of Sun and Earth to spacecraft squared, both distances to the forth power, frequency and frequency drift of the Pioneer transmissions) it seems miraculous that the calculation of the energy density $\rho^*$ in deep space differs by less than 1 % of the value predicted by Curé on the basis of a completely different phenomenon: starlight deflection by the Sun.

---

$^{\dagger}$ To 10 digits, although rightmost digits are not significant due to imprecision of $E_D$

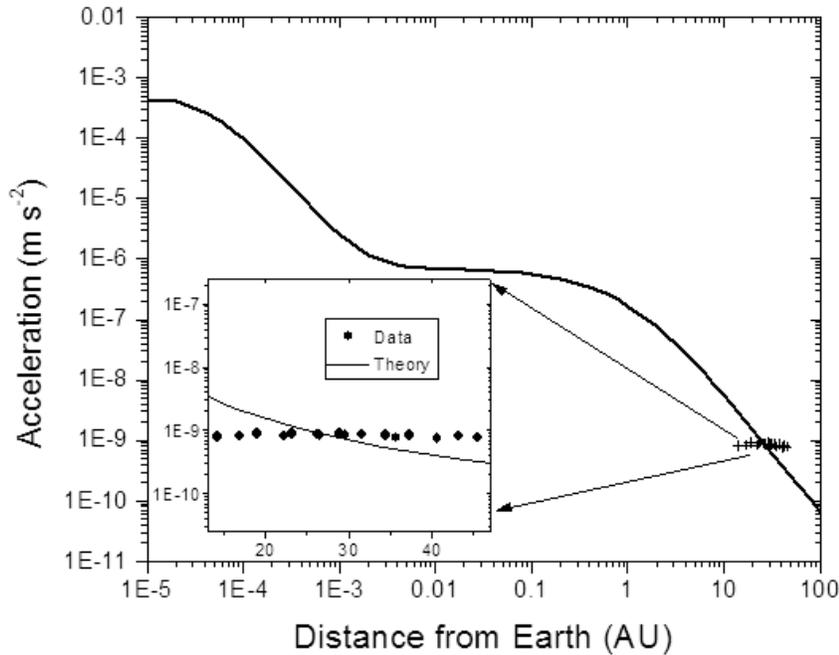

Figure 2. The Pioneer anomalous acceleration predicted with the theory as a function of the distance along a Sun-Earth radial line. (Radius of Earth = 4.26E-05 AU) The experimental data shown, obtained by NASA, was derived from Fig. 7 of Ref. [4].

The puzzling fact that the anomalous acceleration shown by Pioneer is not observed in the planets may be explained: The anomalous acceleration is not real, it is an artefact affecting Doppler measurements of bodies which are in a place where the index of refraction n' ≠ 1 and are in relative acceleration to Earth-bound observers. A Doppler probe on the surface of the planets will show an anomalous acceleration because the energy density of space there is different from the energy density on the surface of Earth. Hence the index of refraction $n'$ on the surface of planets differs from Earth. Table I shows the results of calculating $n'$ with the use of Eq. (7). The values close to 1.0 being caused by the local gravitational energy density being not so different from the surface of the Earth. Values of $n'$ above one indicate that a Doppler probe would show an anomalous acceleration in the direction opposite to the Sun and would be equal to the factor $(1-n')$ times the real relative acceleration of the planet.

Table I. Values of the index of refraction $n'$ in the surface of the planets and the Moon. The value of $\rho^* = 1.09429 \times 10^{15}$ Joule/m$^3$ was used in evaluating $n'$ with equation (7).

| Planet | Mercury | Venus | Earth | Mars | Jupiter |
|---|---|---|---|---|---|
| $n'$ | 0.99997382 | 0.99999527 | 1.00000000 | 0.99997758 | 1.00014145 |

| Planet | Saturn | Uranus | Neptune | Pluto | Moon |
|---|---|---|---|---|---|
| $n'$ | 1.00000349 | 0.99999524 | 1.00000737 | 0.99997385 | 0.99997454 |

## 8.- Conclusion

We find a neo-Newtonian explanation of the Pioneer anomaly. This is done with the Curé [10, p. 173] hypothesis that the speed of light at a site depends on the local space energy density predicted by Newton's universal law of gravitation. With this hypothesis we have been able to deduce in a simple manner the empirically observed phenomenon of the Pioneer anomaly qualitatively and quantitatively. Additionally with the theory developed we are able to calculate the energy density of space produced by the rest of the Universe in the neighbourhood of the Sun. The value obtained ($1.0838. \times 10^{15}$ Joule/m$^3$) coincides very closely with a value ($1.09429 \times 10^{15}$ Joule/m$^3$) deduced by J. C. Curé [10, p. 279] on the basis of the empirical measurement of light bending by the Sun observed during solar eclipses.

The anomalous acceleration does not exist. Pioneer 10 and 11 as well as Galileo and Ulysses spacecraft are moving according to Newton's universal law of gravitation or according to Einstein's General Theory of Relativity which coincide in this respect. The anomaly is found to be due to the effect on the Doppler signals by the index of refraction of space, which is to say the variation of the speed of light due to the energy density of space predicted by the Curé hypothesis.

For further verification of the Curé hypothesis we suggest careful analysis of measurements done on the Pioneer spacecraft in the early stages of the flight, from launch to about 20 AU. Fortunately there are plans at JPL, motorized by S. G. Turyshev, to reanalyze all the data taken of the Pioneer missions, which have now been preserved. [8]

NASA's careful measurements and the Curé hypothesis that the speed of light at a site depends on the local space energy density which explain it have profound implications for physics and cosmology. A lot of other astronomical data needs to be examined in this context. Its acceptance on the basis of the evidence supplied by an explanation of the Pioneer anomaly and the light bending by the Sun obliges a careful revision of the interpretation of data used by Hubble to derive the hypothesis of the expansion of the universe and all the theoretical predictions which follow.

**Acknowledgements**

I would like to thank my colleagues Haydn Barros, Imre Mikoss and Guillermo Chasín for helpful discussions during the development of this work and Gabriel Bernasconi for recent literature on the Pioneer anomaly. Also thank Simon E. Greaves for independent calculation of the numerical values and to Jorge C. Curé for pointing out that the Pioneer anomaly could be explained on the basis of work in his book.